\documentclass[conference]{IEEEtran}
\usepackage{cite}
\usepackage{amsmath,amssymb,amsfonts}
\usepackage{algorithmic}
\usepackage{graphicx}
\usepackage{textcomp}

\def\BibTeX{{\rm B\kern-.05em{\sc i\kern-.025em b}\kern-.08em
    T\kern-.1667em\lower.7ex\hbox{E}\kern-.125emX}}

\title{K-D Bonsai: ISA-Extensions to Compress \\K-D Trees for Autonomous Driving Tasks}
\author{\IEEEauthorblockN{Pedro H. E. Becker, José María Arnau, Antonio González}
\IEEEauthorblockA{\textit{Department of Computer Architecture} \\
\textit{Universitat Politècnica de Catalunya}\\
\{pedro, jarnau, antonio\}@ac.upc.edu}
}
\usepackage{balance} % for balance command
\usepackage{booktabs}
\usepackage{multirow}
\usepackage{tabularx}
\usepackage{tabularray}
\UseTblrLibrary{booktabs}
\usepackage{lipsum}  
\usepackage[acronym,nonumberlist]{glossaries}
\newacronym{AD}{AD}{Autonomous Driving}
\newacronym{AI}{AI}{Artificial Intelligence}
\newacronym[longplural={Application Specific Integrated Circuits}]{ASIC}{ASIC}{Application Specific Integrated Circuit}
\newacronym[longplural={Application-Specific Instruction Set Processors}]{ASIP}{ASIP}{Application-Specific Instruction Set Processor}
\newacronym[longplural=Autonomous Vehicles]{AV}{AV}{Autonomous Vehicle}
\newacronym[longplural={Basic Blocks}]{BB}{BB}{Basic Block}
\newacronym[longplural={Block Random Access Memories}]{BRAM}{BRAM}{Block Random Access Memories}
\newacronym[longplural={Configurable Logic Blocks}]{CLB}{CLB}{Configurable Logic Block}
\newacronym[longplural={Coarse-Grained Reconfigurable Arrays}]{CGRA}{CGRA}{Coarse-Grained Reconfigurable Array}
\newacronym{CMOS}{CMOS}{Complementary Metal–Oxide–Semiconductor}
\newacronym[longplural={Convolutional Neural Networks}]{CNN}{CNN}{Convolutional Neural Network}
\newacronym{DMR}{DMR}{Dual Modular Redundancy}
\newacronym[longplural={Deep Neural Networks}]{DNN}{DNN}{Deep Neural Network}
\newacronym{DSE}{DSE}{Design Space Exploration}
\newacronym{DVFS}{DVFS}{Dynamic Voltage and Frequency Scaling}
\newacronym{DVS}{DVS}{Dynamic Voltage Scaling}
\newacronym{ECC}{ECC}{Error Correction Code}
\newacronym{EDP}{EDP}{Energy-Delay Product}
\newacronym{FP}{FP}{Floating-Point}
\newacronym[longplural={Field-Programmable Gate Arrays}]{FPGA}{FPGA}{Field-Programmable Gate Array}
\newacronym[longplural={Floating-Point Units}]{FPU}{FPU}{Floating-Point Unit}
\newacronym[longplural={Functional Units}]{FU}{FU}{Functional Unit}
\newacronym{GNSS}{GNSS}{Global Navigation Satellite System}
\newacronym[longplural={General Purpose Processors}]{GPP}{GPP}{General Purpose Processor}
\newacronym[longplural=Graphics Processing Units]{GPU}{GPU}{Graphics Processing Unit}
\newacronym{HD}{HD}{High-Definition}
\newacronym{HLS}{HLS}{High-Level Synthesis}
\newacronym{HPC}{HPC}{High-Performance Computing}
\newacronym{ILP}{ILP}{Instruction-Level Parallelism}
\newacronym{IO}{IO}{Input/Output}
\newacronym[longplural={Input/Output Blocks}]{IOB}{IOB}{Input/Output Block}
\newacronym{IP}{IP}{Intellectual Property}
\newacronym{IPC}{IPC}{Instructions per Cycle}
\newacronym{ISA}{ISA}{Instruction Set Architecture}
\newacronym{IMU}{IMU}{Inertial Measurement Unit}
\newacronym{LiDAR}{LiDAR}{Light Imaging Detection and Ranging}
\newacronym[longplural=Least Significant Bits]{LSB}{LSB}{Least Significant Bit}
\newacronym[longplural={Look-Up Tables}]{LUT}{LUT}{Look-Up Table}
\newacronym[longplural={Multiprocessor Systems on Chip}]{MPSoC}{MPSoC}{Multiprocessor Systems on Chip}
\newacronym[longplural={Neural Networks}]{NN}{NN}{Neural Network}
\newacronym[longplural={No-Operations}]{NOP}{NOP}{No-Operation}
\newacronym{OoO}{OoO}{Out-of-Order}
\newacronym{PC}{PC}{Program Counter}
\newacronym{PCL}{PCL}{Point-Cloud Library}
\newacronym{PG}{PG}{Power Gating}
\newacronym{ROB}{ROB}{Re-order Buffer}
\newacronym{ROS}{ROS}{Robot Operating System}
\newacronym{RTL}{RTL}{Register-Transfer Level}
\newacronym[longplural=Self-Driving Cars]{SDC}{SDC}{Self-Driving Car}
\newacronym[longplural=Self-Driving Vehicles]{SDV}{SDV}{Self-Driving Vehicle}
\newacronym{TDP}{TDP}{Thermal Design Power}
\newacronym{VLIW}{VLIW}{Very Long Instruction Word}
\newacronym{HDL}{HDL}{Hardware Description Language} % to load acronyms file and use them with gls and glspl
\usepackage[export]{adjustbox}
\usepackage[dvipsnames]{xcolor}
\usepackage[caption=false]{subfig}
\usepackage{float}

\usepackage{mathptmx} % This is Times font
\usepackage{bm}

\usepackage{fancyhdr}
\usepackage[normalem]{ulem}
\usepackage[hyphens]{url}
\usepackage[final]{microtype}
\usepackage{wrapfig}
\usepackage[bookmarks=true,breaklinks=true,colorlinks,citecolor=blue,linkcolor=blue,urlcolor=blue]{hyperref}
\usepackage{soul}
\soulregister\cite7
\soulregister\ref7
\soulregister\gls7
\soulregister\glspl7

\newcommand\bluecolor[1]{\textcolor{blue}{#1}}
\newcommand*\rot{\rotatebox{90}}

\DeclareRobustCommand{\rebuttal}[1]{{\sethlcolor{SpringGreen}\hl{#1}}}
\soulregister{\rebuttal}{1}

\newcolumntype{b}{X}
\newcolumntype{s}{>{\hsize=.5\hsize}X}          

\begin{document}
\maketitle
\thispagestyle{plain}
\pagestyle{plain}

%%%%%% -- PAPER CONTENT STARTS-- %%%%%%%%
\begin{abstract} 
Autonomous Driving (AD) systems extensively manipulate 3D point clouds for object detection and vehicle localization. Thereby, efficient processing of 3D point clouds is crucial in these systems.  In this work we propose \textit{K-D Bonsai}, a technique to cut down memory usage during \textit{radius search}, a critical building block of point cloud processing. \textit{K-D Bonsai} exploits value similarity in the data structure that holds the point cloud (a k-d tree) to compress the data in memory. \textit{K-D Bonsai} further compresses the data using a reduced floating-point representation, exploiting the physically limited range of point cloud values. For easy integration into nowadays systems, we implement \textit{K-D Bonsai} through \textit{Bonsai-extensions}, a small set of new CPU instructions to compress, decompress, and operate on points. To maintain baseline safety levels, we carefully craft the \textit{Bonsai-extensions} to detect precision loss due to compression, allowing re-computation in full precision to take place if necessary. Therefore, \textit{K-D Bonsai} reduces data movement, improving performance and energy efficiency, while guaranteeing baseline accuracy and programmability. We evaluate \textit{K-D Bonsai} over the euclidean cluster task of Autoware.ai, a state-of-the-art software stack for AD. We achieve an average of 9.26\% improvement in end-to-end latency, 12.19\% in tail latency, and a reduction of 10.84\% in energy consumption. Differently from expensive accelerators proposed in related work, \textit{K-D Bonsai} improves \textit{radius search} with minimal area increase (0.36\%).
\end{abstract}

\section{Introduction}\label{sec:introduction}

\begin{figure}[b]
\begin{minipage}[c]{0.48\linewidth}
\includegraphics[width=\linewidth]{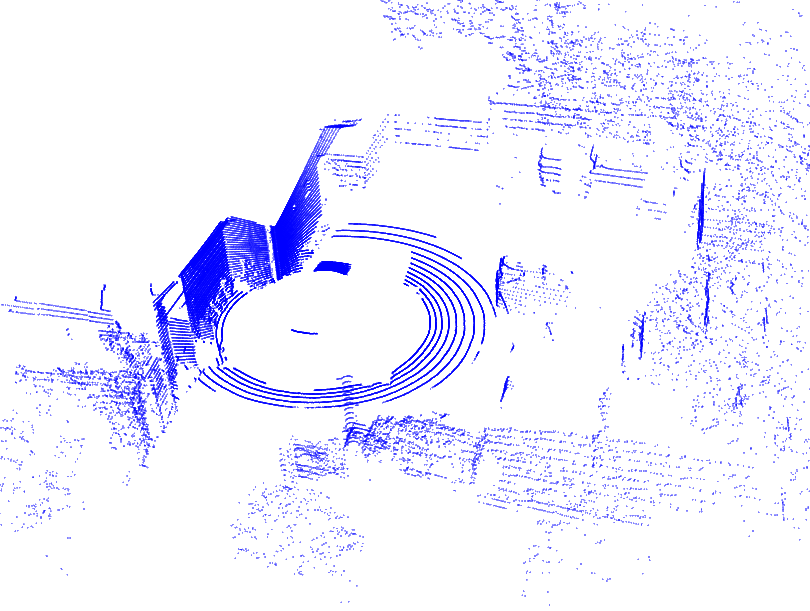}
\caption{A point-cloud obtained with \gls{LiDAR}. Data from \cite{tierIV2020autowareData}.}
\label{fig:pointCloudExample}
\end{minipage}
\hfill
\begin{minipage}[c]{0.48\linewidth}
\includegraphics[width=\linewidth]{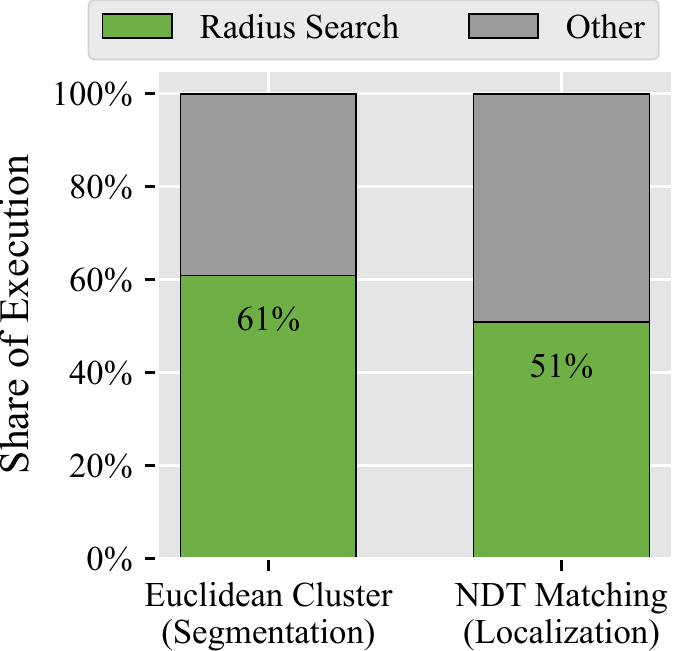}
\caption{\textit{Radius search} execution time share in two Autoware.ai~\cite{autowareGitHub, Kato2015AnOpenApproach} tasks.}
\label{fig:radius_search_share}
\end{minipage}%
\end{figure}

As recent advances in sensors, algorithms, and hardware crystallize the viability of \gls{AD}, concerns shift toward how to make these systems more efficient. In this context, improving hardware support for point cloud manipulation is critical since \glspl{AV} heavily depend on point cloud-based algorithms~\cite{You2020LidarForAutonomousDriving, Kato2015AnOpenApproach, apolloGit}. Point clouds contain a 3D representation of the environment (see Figure \ref{fig:pointCloudExample}), being richer than their 2D counterparts (e.g., images). For this reason, point clouds are suitable for a multitude of tasks, such as object detection, distance measurement, and vehicle localization, which are vital for \gls{AD}.

A crucial point cloud operation performed by \gls{AD} algorithms is \textit{radius search}, whose goal is to return all points within a distance $r$ from a query point $q$ (where $q$ belongs to the point cloud). \textit{Radius search} is used, for example, when clustering nearby points together (to infer shapes and objects around the vehicle)~\cite{klasing2008aClusteringMethod, Rusu2010, nguyen2020FastEuclidean}, or when optimizing the localization estimation of the vehicle~\cite{Besl1992AMethodForRegistration, bieber2003NdtMatching, Magnusson2009ttn}, which are among the most time- and energy-consuming tasks performed by \glspl{AV}~\cite{Becker2020Demystifying, zhao2020DrivingScenario}. In fact, \textit{radius search} accounts for more than half of the execution time of these tasks, as depicted in Figure \ref{fig:radius_search_share} . Likewise, \textit{radius search} is also used in 3D \glspl{CNN} processing~\cite{Lin2021PointAcc, Li2017PointNet++}, in order to fetch neighbors of points to push them together through convolutions.

In this work we propose \textit{K-D Bonsai}, a technique to compress point clouds to reduce data movement during \textit{radius search} execution, improving its performance and energy efficiency. To perform the compression, we first observe that sensors have a physically limited range of operation, defining an upper-bound value for the coordinates of the collected points. For example, the Velodyne HDL-64E~\cite{VelodyneLidar} - a typically employed \gls{LiDAR} sensor - has a maximum operation range in the order of 120 m. This ultimately limits the values of the exponent fields in the \gls{FP} representation of all points in the sensed point cloud. Second, we observe that k-d trees~\cite{bentley1975multidimensional, friedman1977anAlgorithm}, the typical data structure used for efficient point cloud searches (e.g., used by the prominent \gls{PCL}~\cite{rusu2011PCL3DisHere}), intrinsically group points with similar values in the tree leaves. As a consequence, the sign and exponent fields (in IEEE 754 \gls{FP} representation~\cite{IEEEStd754}) are frequently repeated across different points in the same leaf and can be merged. Third, we observe that it is also possible to reduce the size of the mantissa field, and still compute a large percentage of \textit{radius search} without losing radius search precision. More importantly, we show how to cheaply identify any precision loss at run-time, and re-issue full-precision computation to keep baseline accuracy with minimal overheads.

The mechanism is implemented with minor hardware and \gls{ISA} extensions, which we named Bonsai-extensions, into a traditional CPU. We use the Bonsai-extensions to compress the k-d tree during build, and to read and operate over compressed data during traversal. Since it reduces the number of necessary bytes to perform \textit{radius search}, it reduces the number of memory accesses, energy consumption, and execution time. The CPU modifications are punctual, refining over already existing hardware. Moreover, our solution has minimal programmability impact, since improvements are exposed as new CPU instructions, and thus being straightforward to be used in existing applications. This makes it easy to adopt K-D Bonsai in today's systems, in contrast to expensive and hard-to-program out-of-core accelerators.

We validate our idea by extending ARM's AArch64 \gls{ISA} on the gem5 simulator~\cite{gem5_2011, gem5_2020}. We modify the \gls{PCL}~\cite{rusu2011PCL3DisHere} to make use of our new instructions and use it to improve the execution of the euclidean cluster task on Autoware.ai~\cite{Kato2015AnOpenApproach, autowareGitHub} - a state-of-the-art and open-source software stack for \glspl{AV}. We demonstrate how \textit{K-D Bonsai} effectively compress points, reducing the number of necessary load instructions by 23\% and energy consumption by 10.84\%, and improving end-to-end performance by 9.26\% on average and tail-latency by 12.19\%.

In summary, this paper presents the following contributions:

\begin{itemize}
    \item We identify redundancy on bit-fields of \gls{FP} representation in point cloud data stored in k-d trees.
    \item We verify that k-d tree radius search, a critical operation for point cloud-based algorithms in \glspl{AV}, tolerates reduction in format representation.
    \item We derive a mathematical equation to verify whether or not the reduction in format representation could harm the accuracy of the radius search operation, which will trigger re-computation with baseline precision if necessary.
    \item We propose K-D Bonsai, a compression technique to exploit data redundancy and reduction in format representation. K-D Bonsai reduces data movement during \textit{radius search}, improving performance and energy efficiency.
    \item We implement K-D Bonsai as new CPU instructions, namely Bonsai-extensions, demonstrating that our scheme could be easily adopted on next-generation processors for \gls{AD}. We also validate the proposed scheme using a state-of-the-art and open-source software stack for \gls{AD}.
\end{itemize}

The paper is organized as follows. Section~\ref{sec:background} introduces important background concepts such as point clouds, k-d trees, and radius search. Section~\ref{sec:compression} explains how compression can be applied to k-d tree data. Section~\ref{sec:design} discusses the design details, including new instructions and necessary hardware. The results are analyzed in Section~\ref{sec:results}. Finally, we review related work in Section~\ref{sec:related_work} and present final conclusions in Section~\ref{sec:conclusions}.

\section{Background}\label{sec:background}

In this section, we introduce important concepts to contextualize our work. We explain i) point clouds; ii) how they are used by modern \gls{AD} software; iii) the k-d tree data structure, used to search on point cloud data; and iv) the \textit{radius search} operation, used by different \gls{AD} algorithms.

\subsection{Point Cloud for Autonomous Driving}

A point cloud is a set of points in a given coordinate system. In the context of \gls{AD}, point clouds are in the 3D space, where each point has coordinates (\textit{x, y, z}). Point clouds can be obtained with sensors such as \gls{LiDAR}, which sends laser beams around and measures the time for them to reflect back to the sensor~\cite{You2020LidarForAutonomousDriving}. In the absence of sensing noise, each point in a point cloud belongs to a surface in the real world (of a wall, a car, a tree, etc.) within the sensor range, as depicted in Figure \ref{fig:pointCloudExample}.

Given the 3D information held by point clouds, they are commonly used by \gls{AD} systems for perception and localization tasks~\cite{You2020LidarForAutonomousDriving, Kato2015AnOpenApproach}. When used for perception, \gls{LiDAR}-based algorithms serve to understand the surrounding environment. This includes tasks such as object detection and recognition, object tracking, and motion prediction~\cite{You2020LidarForAutonomousDriving, Kato2015AnOpenApproach}. Typical perception algorithms that use point cloud include points clustering~\cite{klasing2008aClusteringMethod, Rusu2010, nguyen2020FastEuclidean} and neural network classification~\cite{Li2017PointNet++, Qi2017, Lang2019}. When used for localization, \gls{LiDAR}-based algorithms try to match the sensed point cloud with a previously existent point cloud map~\cite{Schwarz2010MappingTheWorld}, sometimes referred as \gls{HD} map, in a process known as \textit{registration}~\cite{Besl1992AMethodForRegistration, bieber2003NdtMatching, Magnusson2009ttn}. When the sensed point cloud overlaps an already mapped region, localization can be derived with centimeter-level precision.

\subsection{K-d Tree}

Raw point cloud data obtained from sensors is unorganized. Points are usually pushed back to an array as they are collected by the sensor. For this reason, searching (a common operation across different \gls{LiDAR}-based algorithms) exhaustively on raw point cloud data is prohibitive, especially in the context of \gls{AD} where latency deadlines are strict~\cite{Lin2018ArchitecturalImplications}. The solution is to use a search-friendly data structure, such as a k-d tree \cite{bentley1975multidimensional}.

A k-d tree is a binary tree that allows efficient search in $k$ dimensional data. In this work, we consider the k-d tree implemented by the widely used \gls{PCL}~\cite{rusu2011PCL3DisHere} and FLANN~\cite{Muja2013} libraries, which are adopted by state-of-the-art \gls{AD} software stacks such as Autoware.ai~\cite{autowareGitHub, Kato2015AnOpenApproach} and Baidu Apollo~\cite{apolloSite}. The tree is created as follows.
A root is created when all points are still to be sorted. On each level, starting with the root, the k-d tree selects one coordinate $\textbf{c}$ (among $k$ possible) to split the data. In the 3D case $\textbf{c}$ can be either the $x$, $y$, or $z$ coordinate. The median value in coordinate $\textbf{c}$ across the set of points is found. Points with $\textbf{c}$ value less than the median go to the left sub-tree, and points with $\textbf{c}$ value greater than the median go to the right sub-tree. On each sub-tree, a new splitting coordinate is selected, and the process repeats. A common criterion (e.g., used by \gls{PCL}) is to select the splitting coordinate whose data is more spread out. This helps posterior search traversal to quickly reach the nodes of interest.

Originally, each k-d tree node would hold a point~\cite{bentley1975multidimensional} (e.g., the median point of the splitting coordinate). Later, an optimized k-d tree~\cite{friedman1977anAlgorithm} proposed to store points on leaves only, up to a maximum $m$ number of points per leaf. Whenever a sub-tree contains less than $m$ points, the splitting process stops and the node is defined as a leaf. Restricting points to the leaves reduces the number of examinations while traversing the tree. The optimal number of points per leaf depends on the data and will impact the tree topology (e.g., the tree depth). \gls{PCL} has a default of 15 points per leaf. 

Notwithstanding, during the tree creation, each sub-tree has its bounding box calculated (i.e., the maximum and minimum values in all coordinates that can be found in that sub-tree). The parent node uses this information to hold its distance to each sub-tree, in the splitting coordinate. This will be used when searching, as we explain in the next subsection. Overall, non-leaf nodes on the tree serve to guide the tree traversal during the search, to reach leaves that contain a set of points that fits the search criteria.

\subsection{Radius Search}

The main goal of \textit{radius search} is to return all points within a distance $r$ from a query point $q$. Formally, given a three-dimensional point cloud $\mathcal{P} = \{p_1, p_2, ..., p_N\}$, $p_i \in \mathbb{R}^3$, we want to find the set of neighbor points

$\mathcal{N}(\textbf{q}, r) = \{\textbf{p} \in \mathcal{P} ~|~ dist(\textbf{p}, \textbf{q}) <= r\} $

of a query point $\textbf{q} \in \mathcal{P}, \mathbb{R}^3$, within a distance $r \in \mathbb{R}$. The operation is used, for example, when clustering points from a point cloud, to retrieve the shape of objects in the environment. In that case, \textit{radius search} is successively used to associate nearby points in clusters: e.g., if point $A$ is in the radius of point $B$, and point $B$ is in the radius of point $C$, then $A$, $B$, and $C$ are all parts of the same cluster~\cite{Rusu2010}.

To perform a radius search on a k-d tree one must provide a query point $q$ and the target radius $r$. The tree will be traversed comparing the splitting coordinate value of the current node with the correspondent coordinate of $q$. This comparison gives a best-effort hint of which child sub-tree is closer, and thus more likely to lead to a leaf where points within $r$ can be found. This descending process will lead to the leaf containing $q$ itself (along with other points in that leaf).  When unwinding the tree navigation, the alternative sub-tree (not taken when descending) is also considered. If the distance in the splitting coordinate from $q$ to the sub-tree is smaller than $r$, the sub-tree is visited, and the descending continues. Every time the search finds a leaf, the distance between $q$ and each point $p_i$ on the leaf is calculated. The euclidean distance $d$ is generally used.

\begin{equation}\label{eq:euclidean_distance}
d(q,p_i) = \sqrt{(q_x - p_ix)^2 + (q_y - p_iy)^2 + (q_z - p_iz)^2}
\end{equation}

To avoid performing the square root, a common optimization is to calculate the squared euclidean distance.
\begin{equation}\label{eq:squared_euclidean_distance}
d^2(q,p_i) = (q_x - p_ix)^2 + (q_y - p_iy)^2 + (q_z - p_iz)^2
\end{equation}

Then we can compare $d^2$ with the square radius $r^2$ to classify the point.
\begin{equation}\label{eq:radius_search_classification_squared}
    classification^2(q,p_i) = \begin{cases} in~radius, & \text{if}~d^2 <= r^2 \\ not~in~radius, & \text{if}~d^2 > r^2 \end{cases}
\end{equation}

Whenever $p_i$ is in the radius of $q$, it is added to the radius search result list.

\section{Compressing point clouds on k-d trees for radius search} \label{sec:compression}

In this section, we explain how k-d-trees can be compressed when used for radius search in \gls{AD} tasks, reducing the number of bytes needed to fetch the points during leaf inspection.
We discuss a twofold compression approach that uses both value similarity and a smaller representation. Finally, we discuss the errors introduced (by a smaller representation; value similarity does not introduce any error) and our approach to detecting and correcting them, guaranteeing the baseline accuracy.

\subsection{Compression based on value similarity}\label{subsec:compression_similarity}

\begin{figure}[tb]%
\centering
\subfloat[\centering Nearby points mapped to the same k-d tree leaf node.]{\label{fig:similarity_space_kdtree}\includegraphics[width=0.22\textwidth]{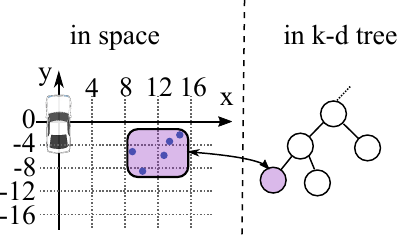}}%
\subfloat[\centering Floating-point fields for each point.]{\label{fig:similarity_sharing_sign_and_exp}\includegraphics[width=0.22\textwidth]{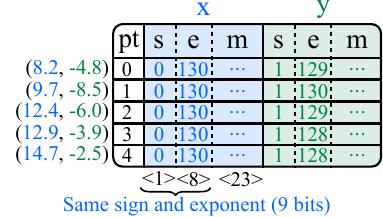}}%
\caption{Nearby points in space are often held by the same k-d tree leaf, creating opportunity to compress data due to value similarity. Particularly, the sign and exponent fields frequently repeat within each point's coordinate.}%
\label{fig:sharing_sign_and_exp}%
\end{figure}

When the k-d tree is built (as explained in section \ref{sec:background}), the point cloud space is subdivided in a way that nearby points end up together in the leaf nodes. Hence, the coordinates of the points are similar to each other. This scenario is illustrated in Figure \ref{fig:sharing_sign_and_exp}, in two dimensions for simplicity.

Figure \ref{fig:similarity_space_kdtree} exemplifies a situation where spatially close points are held by the same k-d tree leaf node. The origin of the coordinate system is in the vehicle (where the \gls{LiDAR} sensor is), and the distance to the points is given in meters. Figure \ref{fig:similarity_sharing_sign_and_exp} lists the coordinates of the points ($x$ and $y$ in this example), exposing their internal \gls{FP} representation (in 32-bit IEEE 754\cite{IEEEStd754}). We depict the sign ($s$), exponent ($e$), and mantissa ($m$) fields of \gls{FP} representation separately. Following the IEEE 754 standard, the stored value is given by the following equation.

\begin{equation}\label{eq:fp_value_from_fields}
    value = -1^{sign}\times{1.mantissa}\times{2^{exponent-bias}}
\end{equation}

When points are close in space, their coordinates are likely to have the same sign (i.e., they all belong to the same quadrant in the coordinate system), and exponent (i.e., values are within the same power of 2). For example, all points in Figure \ref{fig:sharing_sign_and_exp} have their x coordinate between $8.0$ and $16.0$, hence yielding the \textit{same} exponent field value of $130$\footnotemark[1].

\footnotetext[1]{For 32-bit, the bias is 127, resulting in a final exponent of $130-127 = 3$.}

To check the applicability of this observation, we verified how often \textit{sign and exponent} fields are the same for a given coordinate across all points in a leaf node (as it is the case for coordinate $x$ in Figure \ref{fig:similarity_space_kdtree}). We inspected a set of point clouds spanning more than 37 million points that feed the \textit{euclidean cluster} node in Autoware.ai~\cite{Kato2015AnOpenApproach, autowareGitHub} (details about data-set can be found in Section \ref{sec:results}). We identified that 78\% of leaf nodes have the same \textit{exponent} and \textit{sign} for the $x$ coordinate, and 83\% for the $y$ coordinate.

Therefore, value similarity in internal fields of \gls{FP} representation of point clouds is \textit{very} common and a suitable compression source for k-d tree data. If the sign and exponent are the same in a coordinate across all points in a leaf, we can store them only once, and reconstruct the values inside the CPU, only when computation takes place (details in Section \ref{sec:design}).

\subsection{Compression via a smaller representation} \label{subsec:compression_smaller_rep}

Compressing the \textit{sign} and \textit{exponent} of \gls{FP} representation fields (Section \ref{subsec:compression_similarity}) yields a maximum compression ratio of 9 out of 32 bits per coordinate when 32-bit is used - the default in Autoware.ai and \gls{PCL}, and the \textit{baseline} considered in this work. 
To improve the compression ratio further, we need to work over the remaining 23 bits of the \gls{FP} representation which belongs to the \textit{mantissa}.

The problem here is that the \textit{mantissa} field hardly repeats across the points in a leaf. Therefore, compression due to value similarity will not be fruitful for the \textit{mantissa} bits. We can, however, reduce the size of the \gls{FP} representation at the cost of precision. Table \ref{tab:smaller_rep_floating_point} depicts the error in classification (Eq. \ref{eq:radius_search_classification_squared}) using different \gls{FP} formats with less than 32-bits. We use the same set of point clouds as in Section \ref{subsec:compression_similarity}. We experimented with two common 16-bit \gls{FP} representations: \textit{IEEE-754 16-bit} (IEEE half-precision format~\cite{IEEEStd754}), the \textit{bfloat 16} (used for machine learning applications, and e.g., supported by CUDA~\cite{nvidia2022CudaMathAPI}); and also a custom 24-bit representation, for a midway reference in our comparison.

% Please add the following required packages to your document preamble:
% \usepackage{booktabs}
% \usepackage{multirow}
% \usepackage{graphicx}
\begin{table}[tb]
\caption{Classification error of radius search for euclidean clustering using smaller floating-point representations.}
\label{tab:smaller_rep_floating_point}
\resizebox{0.47\textwidth}{!}{%
\begin{tabular}{@{}lcccl@{}}
\toprule
\multirow{2}{*}{} & \multicolumn{3}{c}{\# of bits} & \multirow{2}{*}{Misclassified points} \\ \cmidrule(lr){2-4}
                  & Sign   & Exponent  & Mantissa  &                                       \\ \midrule
IEEE-754 32-bits     & 1      & 8         & 23        & 0\%  (baseline)                    \\
IEEE-754 16-bits     & 1      & 5         & 10        & 0.076\%                               \\
bfloat 16         & 1      & 8         & 7         & 0.61\%                                \\
Custom float 24   & 1      & 5         & 18        & 0.0003\%                              \\ \bottomrule
\end{tabular}%
}
\end{table}

Overall, we found that both 16-bit and 24-bit \gls{FP} representations yield less than 1\% classification error. This is a good indication that reducing the representation can be effective for compression, introducing few mistakes. Notice that for \textit{IEEE-754 16-bit} and the Custom float (24 bits) representations the \textit{exponent} field size is also reduced, affecting the \textit{range} of representable numbers. However, point cloud data obtained from sensors such as \gls{LiDAR} have limited range. For example, the Velodyne HDL-64E~\cite{VelodyneLidar} (a typically employed \gls{LiDAR} sensor) has a maximum cover range of 120 m. Indeed, none of the errors depicted in Table \ref{tab:smaller_rep_floating_point} are due to the lack of range to represent numbers. Hence, reducing \textit{exponent} bits in our case is not a problem, but something to take advantage of\footnotemark[2].

\footnotetext[2]{Lack of range representation due to fewer exponent bits could be a problem when the coordinate system of the point cloud does not have the origin on the sensor itself and is, otherwise, far away. For example, when point cloud maps~\cite{seif2016hdMap, Schwarz2010MappingTheWorld} are created, several point clouds are combined to represent a region. Hence, points can be more distant to the origin than the sensor range. A possible solution for this case is to translate the origin to a more convenient position. This could be done offline or when the map of the region is loaded.}

Going further, we evaluate the involved trade-offs of the different representations to select a good fit for our compression scheme. We noticed that \textit{IEEE-754 16-bit} has the same size as \textit{bfloat}, but balances better the use of exponent bits (for range) and mantissa bits (for precision), being more accurate by an order of magnitude. Also, the 8 extra bits in our Custom (24 bits) float for increased precision do not pay off since the 16-bit formats already hold decent ($<$1\% error) accuracy. Finally, the \textit{IEEE-754 16-bit} is already partially supported by nowadays CPUs (e.g., for storage on ARM~\cite{ARM2021}) hence being less intrusive on existing architectures than a new custom format. For these reasons, we choose the \textit{IEEE-754 16-bit} to represent the points of k-d tree leaves, and over that apply compression due to value similarity (Section \ref{subsec:compression_similarity}).

Our main conclusions about using a smaller representation in k-d tree radius search are two-fold: i) the \textit{mantissa} bits can be reduced with low accuracy loss; ii) \gls{AD} algorithms consume points that are near the vehicle, hence the \textit{exponent} bits can be reduced and still represent the point cloud values.

\subsection{How to keep accuracy despite a smaller representation}\label{subsec:keep_accuracy}

So far, we have discussed two different ways to reduce the size of points searched by k-d trees, with the side effect of introducing classification errors. However, since \gls{AD} systems are safety-critical, introducing mistakes is not desirable~\cite{hussain2019AutonomousCars} and pose consequences which are hard to test~\cite{koopman2016challengesInAVTesting}. Hence, we propose an approach to detect possible mistakes in classification, and re-compute them with baseline accuracy. For this, we assume to have access to both the original points and the compressed points. The idea is to use the compressed points, alleviating memory usage, and exceptionally lookup for the original 32-bit values if a possible misclassification is detected.

Let $B$ be a number in \textit{32-bits IEEE-754} format that we want to represent in the \textit{16-bit IEEE-754} format, at the cost of an error ${\delta}B$ associated with the loss of precision. Let $B'$ denote the resulting value of $B$ in 16-bit representation.

\begin{equation}\label{eq:single_num_reduced_precision}
    B' = B + {\delta}B
\end{equation}

For the default rounding mode in the IEEE-754 Standard, the \glspl{LSB} of the mantissa are dropped, and the resulting number is rounded up or down, towards the nearest number. For values whose exponent can be stored equally in both representations (our case, see Section \ref{subsec:compression_smaller_rep}), the rounding in the mantissa is the single source of error. In this case, the $11^{th}$ to $23^{rd}$ \textit{mantissa} bits will be used to round the number to its nearest value, adjusting the $10^{th}$ bit of the 16-bit resultant number. 

Since we can round up or down to the nearest number, the \textit{maximum mantissa error} will be half the value of the $10{th}$ bit, while the \textit{maximum value error} will also depend on the exponent, since $2^{exponent-bias}$ multiplies the mantissa to form the \gls{FP} number (Eq. \ref{eq:fp_value_from_fields}). In these conditions, the maximum error $\delta$ for rounding a number $B$ when converting it from 32-bit to 16-bit IEEE-754 \gls{FP} is given by:

\begin{equation}\label{eq:max_delta}
    max({\delta}B) = 2^{exponent-bias}\times{\frac{2^{-10}}{2}} = 2^{exponent-bias}\times{2^{-11}}
\end{equation}

The takeaway here is that using only the exponent one can infer the maximum rounding error. Thus, with $B'$ at hand, there is no need to lookup $B$, as the exponent value is representable in both $B'$ and $B$ according to our assumptions.

Now, let's proceed to find the error in the squared difference between a value $A$, in 32-bit, and a value $B'$, in 16-bit. We start looking at the subtraction, applying Eq. \ref{eq:single_num_reduced_precision}.

\begin{equation}\label{eq:error_subtraction}
    \begin{array}{lclcl}
        A - B' & =  & (A) - (B + {\delta}B) & = & (A - B) -  {\delta}B
    \end{array}
\end{equation}
Where $-{\delta}B$ is the associated error. We can proceed and evaluate the error for the square operation $(A-B')^2$ applying Eq. \ref{eq:single_num_reduced_precision}, Eq. \ref{eq:error_subtraction}, and Newton's binomial theorem.
\begin{equation}\label{eq:error_square}
    \begin{array}{lcl}
        (A-B')^2 & =  & [(A-B) -  {\delta}B]^2 \\
        & = & (A-B)^2 - 2(A-B){\delta}B + {\delta}B^2 \\
        & = & (A-B)^2 - 2[A-(B' - {\delta}B)]{\delta}B + {\delta}B^2 \\
        & = & (A-B)^2 - 2(A-B' + {\delta}B){\delta}B + {\delta}B^2 \\
        & = & (A-B)^2 - 2[(A-B'){\delta}B + {\delta}B^2] + {\delta}B^2 \\ 
        & = & (A-B)^2 - 2(A-B'){\delta}B -2{\delta}B^2 + {\delta}B^2 \\
        & = & (A-B)^2 - 2(A-B'){\delta}B -{\delta}B^2
    \end{array}
\end{equation}
Where $- 2(A-B'){\delta}B -{\delta}B^2$ is the associated error of the square of the differences operation ($\epsilon_{sd}$). Notice that ${\delta}B$ can be either positive or negative, depending if the number was rounded up or down. At run-time, however, we will not know which case it was because that would require fetching and inspecting the \glspl{LSB} of the original value, which we are trying to avoid. Instead, we can be pessimistic and calculate the \textit{worst case} magnitude of $\epsilon_{sd}$, using the $max({\delta}B)$  (Eq. \ref{eq:max_delta}) instead of ${\delta}B$. % avoid fetching the original 32-bit value.

\begin{equation}\label{eq:error_magnitude}
    \begin{array}{lcl}
        max(\epsilon_{sd})& = & 2\cdot |A - B'| \cdot |max({\delta}B)| + max({\delta}B)^2
    \end{array}
\end{equation}

Again, notice that the $max({\delta}B)$ and $max({\delta}B)^2$ can be directly obtained with the exponent of $B'$. Finally, we can compute the approximate square differences of form $(A-B')^2$ for each coordinate, and sum to get the approximate euclidean distance squared $d'^2$.
\begin{equation}\label{eq:approx_squared_euclidean_distance}
d'^2(q,p'_i) = (q_x - p'_ix)^2 + (q_y - p'_iy)^2 + (q_z - p'_iz)^2
\end{equation}

Likewise, we can sum the maximum error of the squared differences in each coordinate and get a total error $T\epsilon_{sd}$

\begin{equation}\label{eq:final_error}
T\epsilon_{sd}= max(\epsilon_{sd})_x + max(\epsilon_{sd})_y + max(\epsilon_{sd})_z
\end{equation}

We can finally use Eqs. \ref{eq:approx_squared_euclidean_distance} and \ref{eq:final_error} to perform the classification (with $p'_i$ instead of $p_i$).
\begin{equation}\label{eq:radius_search_classification_with_error}
    classification'^{2}(q,p'_i) = \begin{cases} in~radius, & \text{if}~d'^2 <= r^2 - T\epsilon_{sd} \\ not~in~radius, & \text{if}~d'^2 > r^2 + T\epsilon_{sd} \\
    use~Eq.~\ref{eq:radius_search_classification_squared}, & \text{otherwise} \end{cases}
\end{equation}

\begin{figure}[bt]
\centering
\includegraphics[width=0.38\textwidth]{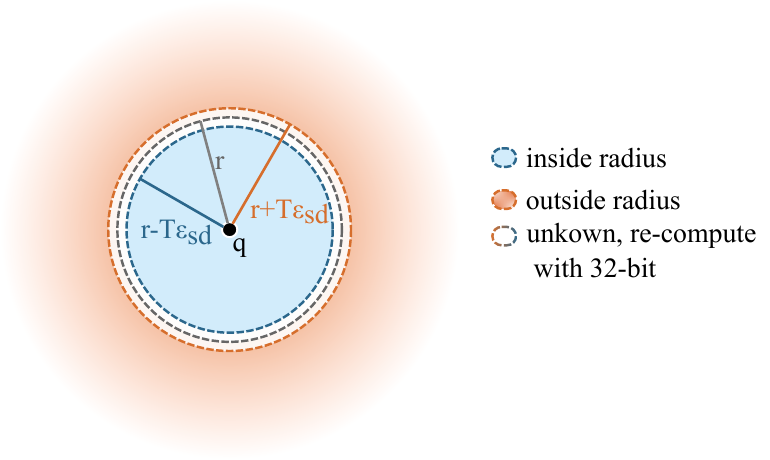}
\caption{Visual representation of Equation \ref{eq:radius_search_classification_with_error}.}
\label{fig:radius_classification}
\end{figure}

In other words, we can use the worst-case error $T\epsilon_{sd}$ to confirm the correctness of the classification with $p'_i$. We do so by defining a shell around $r^2$ with values $r^2 - T\epsilon_{sd}$ and $r^2 + T\epsilon_{sd}$, as depicted in Figure \ref{fig:radius_classification}. Whenever $d'^2$ falls outside the shell, the classification is the same as the baseline, computed by Eq. \ref{eq:radius_search_classification_squared}. For instance, a point inside the radius but outside the shell cannot be outside the radius even if we add $T\epsilon_{sd}$ to $d'^2$.  On the other hand, when $d'^2$ falls inside the shell, the error could be large enough to change the classification, and cannot be guaranteed to be the same as the baseline. In this case, we propose to fetch the original point $p_i$, and re-do the classification with the full-precision, using Eq. \ref{eq:radius_search_classification_squared}.

\section{Proposed Design} \label{sec:design}

In this section, we motivate and explain the design decisions of \textit{K-D Bonsai}. We explain the hardware structures and how to use them through new instructions, the \textit{Bonsai-extensions}.

\subsection{Hardware support for k-d tree compression}
After deriving a compression scheme (Section \ref{sec:compression}), hereby referred to as \textit{K-D Bonsai}, it is of our interest to use it in tasks that perform \textit{radius search}. A naïve approach would be to (de)compress points with a software-only solution. However, iteratively inspecting and re-ordering bits in software slows down \textit{radius search} in the order of 7$\times$ (data-set and experimentation platform in Section \ref{subsec:results_methodology}), undermining the compression benefits. Alternatively, it is possible to add hardware to support \textit{K-D Bonsai} effectively.

Two main options arise to implement \textit{K-D Bonsai} in hardware: i) with an out-of-core accelerator; or ii) in the CPU through \gls{ISA}-extensions. In this work, we stand with the latter as we justify next. First, the CPU would have to transfer data in and out to communicate with the accelerator. However, the leaf processing done by K-D Bonsai is a fine grain task, requiring only a handful of cycles to complete (implementation details in Section \ref{subsec:changing_cpu}). Thus, using proper hardware inside the CPU to perform (de)compression and classify points avoids communication costs~\cite{Shao2016accDesign}. Alternatively, (de)compression operations could be coalesced to amortize communication costs. However, accelerators are likely to be more expensive (see Section \ref{sec:related_work}). At the same time, leaf processing is only a fraction of the point cloud handling, limiting the maximum performance improvement (Ahmdal's law), and jeopardizing accelerator adoption. Nevertheless, industry favors less experimental approaches to accelerate tasks in their real-life solutions, rarely employing accelerators~\cite{Peccerillo2022SurveyAccelerators}.

On the other hand, while new instructions yield more conservative performance gains, they are a much simpler solution from the hardware standpoint. Additionally, \gls{ISA}-extensions are easier to integrate and to program, facilitating \textit{K-D Bonsai} implementation in existing platforms. For example, ARM releases new (sometimes optional) \gls{ISA}-extensions yearly~\cite{arm2019understandingARMv8x}. Also, some ARM processors support to-be-defined custom instructions~\cite{Choquin2020}. Both alternatives exemplify the use of \gls{ISA}-extensions to specialize CPUs for relevant scenarios, such as \gls{AD}. Support for custom instructions is also a key feature of the RISC-V \gls{ISA}~\cite{riscvISA}. This set of reasons motivates us to propose specific instructions in the CPU to implement \textit{K-D Bonsai} effectively.

\subsection{Changing the CPU}\label{subsec:changing_cpu}

A main advantage of the ideas discussed in Section \ref{sec:compression} is how easily and cheaply they can be carried out in the hardware. Indeed, the set of new functionalities required is small: i) we need to compress the data; ii) decompress the data; and iii) support computation of the squared differences (and associated error) in the form $(A-B')^2$ (Eq. \ref{eq:error_square}).

\begin{figure}[bt]
\centering
\includegraphics[width=0.35\textwidth]{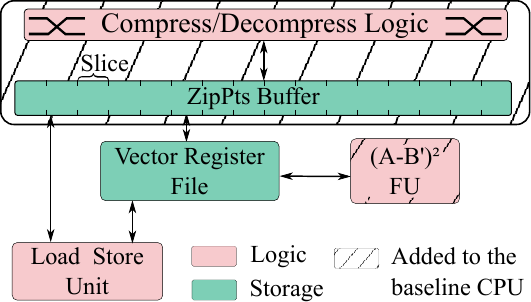}
\caption{The new components added to the baseline CPU and how they interact with pre-existing ones.}
\label{fig:new_cpu}
\vspace{-10pt}
\end{figure}

Figure \ref{fig:new_cpu} depicts the two components that we add to the CPU, and how they interact with the existing hardware. The first added component we discuss is the Compression/Decompression unit, at the top of the figure. The unit is divided into two parts: a buffer, named ZipPts Buffer, and a Compress/Decompress Logic.

%figures/new_cpu.pdf

\textbf{ZipPts Buffer.} The ZipPts Buffer is designed to hold both compressed and uncompressed 16-bit points, being the source and destination operand for compression and decompression operations. In our implementation, we restrict the ZipPts Buffer size to hold a maximum of 16 points (the number of points per leaf in the \gls{PCL} is 15 by default). We also reserve space for 3 bits in the buffer, to encode whether $x$, $y$, and $z$ coordinates are compressed.

The buffer has two 128-bit ports to interface with the Vector Register File and one 128-bit port to interface with the Load Store Unit. Hence, data is exchanged in chunks of 128-bit, which we refer to as a \textit{ZipPts Buffer slice}. When less than 128-bits must be transferred (e.g., the last chunk of a compressed data), we pad data with zeroes.  The width of the ports equals the ones that already exist in our baseline CPU (see Section \ref{sec:results}), for example in the Vector Register File. Hence, we can load and store data from/to memory directly to the ZipPts Buffer. In summary, we can load points into the buffer to be compressed, store the compressed data back in the memory, and load compressed data to be decompressed. Also, we can write values from the ZipPts Buffer into the Register File, exposing them to the \glspl{FU}. The ZipPts Buffer is tightly coupled with the Compress/Decompress Logic, which is responsible for re-arranging the data bits, discussed as follows.

\begin{figure*}[bt]
\centering
\includegraphics[width=\textwidth]{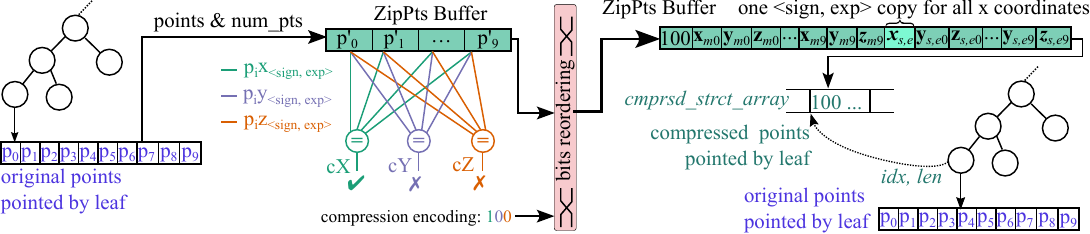}
\caption{Compressing points in a leaf node. Load the points into the ZipPts Buffer. Find coordinates with same $<sign, exponent>$ pairs, e.g., $x$ coordinate (setting $cX$ to 1). Reorder bits in the ZipPts Buffer and set the compression encoding. Store compressed data in the memory (\textit{cmprsd\_strct\_array}), adding a reference to it in the leaf node for future look-up.}
\label{fig:compression_workflow}
\end{figure*}

\textbf{Compress/Decompress Logic.} This unit re-arranges the data in the ZipPts Buffer compressing and decompressing points from a k-d tree leaf. In both cases, the number of points must be provided to the logic. During compression, this unit reads and compares the tuple $<sign, exponent>$ on each coordinate of the points in the ZipPts Buffer, see Figure \ref{fig:compression_workflow}. If they are the same across all points, only one \textit{copy} of $<sign, exponent>$ will appear in the resulting compressed data. Each coordinate has a compression bit flag ($cX$, $cY$, $cZ$) to indicate whether or not its $<sign, exponent>$ is compressed. During decompression, this unit reads the compression bit flags, re-organizing the data and re-creating the multiple instances of the single copy of $<sign, exponent>$ across all values.

To exemplify, Figure \ref{fig:compression_workflow} details the compression flow and the organization of the compressed data. First, the mantissa values are directly bypassed to the buffer as they are not compressed. Then, the compressed tuples of $<sign, exponent>$ are placed in the ZipPts Buffer, followed by the remaining non-compressed tuples of $<sign, exponent>$. The three compression bits are placed at the very beginning of the buffer. 

\begin{figure}[bt]
\centering
\includegraphics[width=0.35\textwidth]{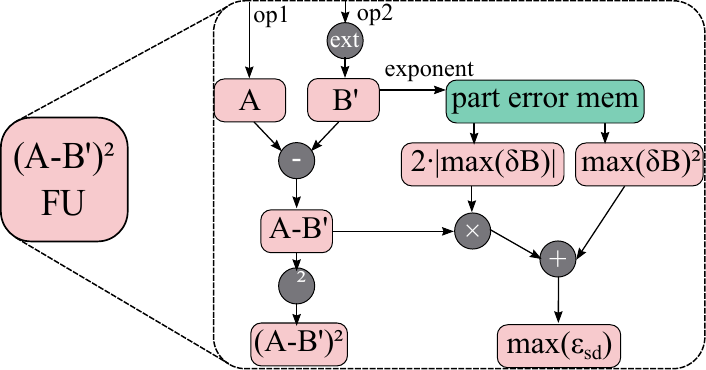}
\caption{Details of the FU for square of the difference with error computation.}
\label{fig:sq_diff_fu}
\end{figure}

\textbf{Approximate Square of Differences Functional Unit.} When compressed points are fetched from memory and decompressed into 16-bit values, they can be moved from the ZipPts Buffer to the Vector Register File. At this point, the \gls{FU} for the square difference with error computation can take place. The unit implements Eq. \ref{eq:error_square}, and can be used successive times (for each coordinate) to compute Eq. \ref{eq:approx_squared_euclidean_distance} and \ref{eq:final_error} to perform the classification. Figure \ref{fig:sq_diff_fu} details the internal scheme of the \gls{FU}. It has two input operands, A is a 32-bit value (e.g., a coordinate of the query point), and B' is a 16-bit value (e.g., the same coordinate of one of the points in the leaf), which is then extended to 32-bit (without changing the value of $B'$) so computation takes place in 32-bit hardware, preventing 16-bit errors to be magnified. The square of the differences proceeds with conventional subtraction and square operations.

The calculation of the worst case error ($max(\epsilon_{sd})$, Eq. \ref{eq:error_magnitude}) has more operations than the square of the differences itself. Fortunately, we can take advantage of some observations to simplify its computation. First, since the $max({\delta}B)$ depends only on the exponent of $B'$, and there are only $2^5=32$ possible exponents, we can pre-compute the values of $2\cdot|max({\delta}B)|$ and $|max({\delta}B)|^2$ and store them in a small (32 lines) lookup table. This small table (named \textit{part error mem} in Figure \ref{fig:sq_diff_fu}) is looked up with the exponent of $B'$ in the beginning of the operation. Also, the term $|A - B'|$ computed for the square of the differences can be borrowed to compute the worst-case associated error $max(\epsilon_{sd})$.

\begin{figure}[bt]
\centering
\includegraphics[width=0.4\textwidth]{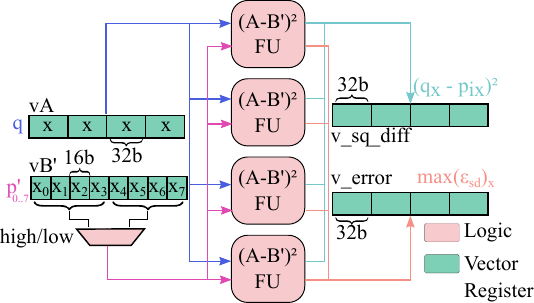}
\caption{Vector square of the differences FUs.}
\label{fig:sq_diff_fu_vector_instruction}
\end{figure}

Since decompression outputs multiple points at once, they are simultaneously available for computation. To leverage this, we instantiate multiple approximate squares of difference \glspl{FU} (Figure \ref{fig:sq_diff_fu_vector_instruction}), to compute them in a vector manner. In each \gls{FU} we compute the square of the differences and the associated error at the same time, each working on a part of the input vectors $vA$ and $vB'$. For the radius search classification, a coordinate of the query $q$ is loaded into all indices of $vA$, while the same coordinate of multiple points is loaded into $vB'$.

\subsection{Software impact}\label{subsec:sw_impact}

Now we discuss how to use the new hardware from the software. We expose the new hardware functionalities mentioned in Section \ref{subsec:changing_cpu} as new CPU instructions. The set of new instructions, which we refer to as \textit{Bonsai-extensions}, is described in Table \ref{tab:instructions}. We divide the \textit{Bonsai-extensions} into three instruction categories: compress, decompress, and computation. Some instructions trigger multiple micro-operations, as we explain together with their usage following.

When the leaf node is created during the k-d tree construction, we can use the compress instructions over the leaf points we have at hand (Figure \ref{fig:compression_workflow}).  For such we have to load the points, one by one, into the ZipPts Buffer using the \textbf{LDSPZPB} instruction. The load converts the original 32-bit into 16-bit before placing the coordinates in the buffer. We can further compress the data in the ZipPts Buffer, looking for sign and exponent sharing, with the \textbf{CPRZPB} instruction. At this point, we have a compressed structure in the ZipPts Buffer and the resulting size in bytes (length). We can proceed and store the compressed data with the \textbf{STZPB} instruction, indicating the amount of \textit{ZiptPts Buffer slices} that must be stored in memory. The decoder will generate one store micro-operation for each \textit{slice}, storing them in consecutive addresses.

In our modified \gls{PCL} code, we create an extra array of bytes, \textit{cmprsd\_strct\_array}, to store the compressed structures consecutively as we visit and compress leaf nodes during the tree construction. Also, we keep track of the starting address and length of the compressed structure placed in the \textit{cmprsd\_strct\_array} in the k-d tree, so that we can fetch the compressed data later, during the radius-search (tree traversal). We use C unions to re-use fields of the tree that are not used on leaf nodes (e.g., the splitting coordinate and distances to children), to store this information. Hence, we hold auxiliary compression information without increasing the size of the k-d tree. In the \gls{PCL} code, we also keep track of the next free index in the array, to be occupied by the next compressed structure.

% Please add the following required packages to your document preamble:
% \usepackage{booktabs}
\begin{table*}[bt]
\footnotesize
\centering
\caption{The Proposed Bonsai-extension Instructions}
\begin{tblr}{X[-1,l,m]X[3,l,h]X[9,halign=j,h]}
\toprule
 & \SetCell[]{c}Instruction & \SetCell[]{c}Description \\
\midrule
\SetCell[r=3]{m,c} \rot{Compress} 
& \textbf{LDSPZPB} r\_index, [r\_addr] 
& \textbf{L}oa\textbf{D} \textbf{S}ingle-float \textbf{P}oint into \textbf{Z}ip\textbf{P}ts \textbf{B}uffer - Loads one 3D point in single-float from address \textit{[r\_addr]}, converts it to 16-bit, and place it on the ZipPts Buffer at position \textit{[r\_index]}.
\\
\cmidrule{2-3}
& \textbf{CPRZPB} r\_size, r\_num\_pts 
& \textbf{C}om\textbf{PR}ess \textbf{Z}ip\textbf{P}ts\textbf{B}uffer - Compress the 16-bit points from the ZipPts Buffer, exploiting the value similarity concept (Section \ref{subsec:compression_similarity}). The number of points is informed in \textit{r\_num\_pts}. The result of the compression is the ZipPts Buffer itself. The size in bytes of the resulting compressed structure is placed in \textit{r\_size}.
\\
\cmidrule{2-3}
& \textbf{STZPB} [r\_addr], \#ZipPtsSlices
& \textbf{ST}ore \textbf{Z}ip\textbf{P}ts\textbf{B}uffer - Stores the ZipPts Buffer in the memory. Due to port size limitations, the ZipPts Buffer will be stored in slices through several store micro-operations (in a total of \textit{\#ZipPtsSlices}).
\\
\cmidrule{1-3}
\rot{Decompress} 
& \textbf{LDDCP} v\_base, r\_num\_pts, [r\_addr], \#ZipPtsSlices
& \textbf{L}oa\textbf{D} \textbf{D}ecompressing \textbf{C}ompressed \textbf{P}oints - Load the compressed structure from memory into the ZipPts Buffer, in slices, through several load micro-operations (in a total of \textit{\#ZipPtsSlices}). Decompress the ZipPts Buffer on itself with one micro-operation. Writes-back the points to vector registers, per coordinate, from \textit{v\_base} up to \textit{v\_base + 5}, with 3 micro-operations. Since two 128-bit registers can hold up to sixteen 16-bit values (enough for one coordinate), we write-back to \textit{six} (two at a time) 128-bit registers to hold forty-eight 16-bit values (enough for three coordinates).
\\
\cmidrule{1-3}
\SetCell[r=2]{m,c} \rot{Computation} 
& \textbf{SQDWEL}  v\_sq\_diff, v\_error, vA, vB'
& \textbf{SQ}uare \textbf{D}ifference \textbf{W}ith \textbf{E}rror \textbf{L}ow part - Performs a vector operation in the form $(A_{i}-B_{i}')^2$ with error calculation (see Eq. \ref{eq:error_square}). The four values in the low part of \textit{vB'} will be extended from 16-bit to 32-bit when pushed in the units (see Figures \ref{fig:sq_diff_fu} and \ref{fig:sq_diff_fu_vector_instruction}). The square difference will be placed in \textit{v\_sq\_diff}, and the associated error in \textit{v\_error}.
\\
\cmidrule{2-3}
& \textbf{SQDWEH}  v\_sq\_diff, v\_error, vA, vB'
& \textbf{SQ}uare \textbf{D}ifference \textbf{W}ith \textbf{E}rror \textbf{H}igh part - Same as SQDWEL, but using the high part of \textit{vB'}.\\
\bottomrule
\end{tblr}
\label{tab:instructions}
\end{table*}

Later, when we do the radius search we can use the \textbf{LDDCP} instruction to load and decompress the compressed structure into registers, whenever we reach a tree leaf. This instruction is broken down by the decoder into a sequence of micro-operations. First it loads the compressed structure into the ZipPts Buffer. For this we need the address and size of the compressed data in the \textit{cmprsd\_strct\_array}, which is kept in the tree leaf, to indicate how many \textit{slices} (chunks of 128-bits) must be brought from memory, starting from the provided address. The decoder use the indicated number of \textit{slices} to generate an equivalent number of load micro-operations from memory to the ZipPts Buffer. Once the whole compressed structure is inside the ZipPts Buffer, a decompression micro-operation takes place, reading the compression encoding and reordering the bits into 16-bit points accordingly. Finally, write-back micro-operations are issued to move the value of the points into the vector register file. In this case, we write back the decompressed points from the ZipPts Buffer into six vector registers. We need two vector registers for each coordinate since each vector register can hold up to eight 16-bit values, and we support up to sixteen 16-bit values per coordinate.

Finally, when we have decompressed the 16-bit values of the coordinates in the vector arrays, we can use the square of the differences \glspl{FU} (Figure \ref{fig:sq_diff_fu_vector_instruction}). For such, we perform instructions \textbf{SQDWEL} and \textbf{SQDWEH}, calculating the square of differences for points with a vector of the query point, for each coordinate. The coordinate values of the query point can be loaded into vector registers using existing vector instructions. Since we have \textit{four} 32-bit lanes in the baseline CPU SIMD unit (ARM NEON, details in Section \ref{subsec:results_methodology}), but \textit{eight} values on each coordinate (16-bit computed in 32-bit in the \glspl{FU}, Figure \ref{fig:sq_diff_fu}), we split the values in two groups of \textit{four} values, the low part and the high part, and compute them one at a time in the four lanes (details in Figure \ref{fig:sq_diff_fu_vector_instruction}). The result, per point index, is available in two vector registers, one holding the calculated square of the differences, and another one with the maximum error ($max(\epsilon_{sd})$). Thereafter, it is possible to accumulate the distances for each index on each coordinate, using existing instructions, and compare it with $r^2$, performing the classification (Eq. \ref{eq:radius_search_classification_with_error}). If the result is inconclusive (inside the white shell in Figure \ref{fig:radius_classification}), one can proceed with the baseline code, i.e., read the 32-bit point and compute the 32-bit distance. This should be rare to guarantee good performance, otherwise compression/decompression will consume time with no real benefit.

Finally, we highlight that, for \gls{AD} tasks, the tree is generally built once for each frame, in the beginning, and then searched multiple times, during the frame processing. This is important because compressing the leaf node points represents an overhead during tree creation. However, the compression benefits will appear during the search, when we load fewer data from the memory. For example, we verified an average of 52 visits for each created leaf node during the radius search for one of the input frames. Thus, the expectation is that loading less data, multiple times, amortizes the initial overhead.

\section{Results} \label{sec:results}

In this section, we explain the evaluation methodology and obtained results for K-D Bonsai.

\subsection{Evaluation Methodology}\label{subsec:results_methodology}

From the software perspective, we rely on Autoware.ai~\cite{Kato2015AnOpenApproach} to experiment with our idea. Autoware is a state-of-the-art and open source software stack for \gls{AD}, built with contributions from both academia and industry companies~\cite{autowareGitHub}. It has several algorithms to perform \gls{AD}, from sensor processing and perception to actuation. In this work, we choose a representative algorithm from Autoware, namely euclidean cluster~\cite{Rusu2010} to verify the benefits of our proposal in k-d tree \textit{radius search}, although other algorithms are also subject to our optimizations (e.g., Autoware's localization algorithm~\cite{Magnusson2009ttn}).

The euclidean cluster algorithm is a vital part of the perception pipeline of Autoware.ai. The algorithm clusters points of a source point cloud, useful for inferring objects' shape, geometry, and distance. Notably, it has been reported by previous works as one of the tasks with higher latency in the Autoware.ai pipeline~\cite{Becker2020Demystifying}. Importantly, the euclidean cluster extensively performs the \textit{radius search} operation to find nearby points that should belong to the same cluster.

We stimulate the euclidean cluster algorithm with a \textit{subset} of point cloud frames from an eight-minute car driving sequence~\cite{tierIV2020autowareData}. Because our cycle-accurate simulator (details next) executes several orders of magnitude slower than real hardware, we used systematic sub-sampling (fixed-size samples equally spaced in time) to select the \textit{subset} of point cloud frames. The idea was inspired by previous work \cite{ardestani2013ESESC} and yields good results if the parameters (interval amount and length) are properly chosen. We experimented with several parameters finally settling on 20 samples of 300 milliseconds each -- adding up to six seconds of real-life data and handling a total of 60 frames. Table \ref{tab:error_subsampling} details sub-sampling errors, evidencing it as a fast and accurate proxy to the code behavior.

% Please add the following required packages to your document preamble:
% \usepackage{booktabs}
\begin{table}[tb]
\centering
\caption{Sub-sampling Error}
\label{tab:error_subsampling}
\begin{tabular}{@{}cccc@{}}
\toprule
\begin{tabular}[c]{@{}c@{}}Mean Standard \\ Error for Latency\end{tabular} & \begin{tabular}[c]{@{}c@{}}IPC Relative\\ Error\end{tabular} & \begin{tabular}[c]{@{}c@{}}L1- D Cache Miss \\ Ratio Difference\end{tabular} & \begin{tabular}[c]{@{}c@{}}Branch Mispred.\\ Difference\end{tabular} \\ \midrule
2.94\% & 4.68\% & 0.10\% & 0.03\% \\ \bottomrule
\end{tabular}
\end{table}

We implemented the Bonsai-extensions (Table \ref{tab:instructions}) in the gem5 simulator~\cite{gem5_2011, gem5_2020}, targeting an \gls{OoO} CPU with the ARM's AArch64 \gls{ISA}. We base our model (see Table \ref{tab:cpu_model}) on the pre-defined \textit{big} CPU in gem5, adjusting parameters such as the frequency to match technology scaling, to replicate an ARM Cortex A72 behavior. Although our solution is \gls{ISA}-agnostic, we used ARM as a representative \gls{ISA} for \gls{AD} (e.g., used by NVIDIA DRIVE~\cite{nvidia2019nvidiaDrive}). We modified the \gls{PCL}~\cite{pointCloudLibrary} version 1.10 and its auxiliary library FLANN~\cite{Muja2013} version 1.9.2, using our instructions during the \textit{radius search}, as explained in Section \ref{subsec:sw_impact}. We did not modify the compiler but instead wrote our instructions directly with byte-code using the \textit{.inst} directive in ARM \textit{asm} inside the library. We expose a Boolean variable in \gls{PCL} so that users can activate the use of the new instructions for \textit{radius search}. When the variable is \textit{true}, the code uses the Bonsai-extensions, otherwise, it uses the baseline code. The search result is the same in both cases.

% Please add the following required packages to your document preamble:
% \usepackage{booktabs}
\begin{table}[b]
\small
\centering
\caption{Baseline CPU Model Used}
\label{tab:cpu_model}
\begin{tblr}{width=0.47\textwidth, colspec={X[1]X[5]}}
\hline
Parameter & Value \\ \midrule
CPU & OoO ARM v8 64-bit @3GHz, Fetch Width: 3, Issue Width: 8, Int Physical Reg.: 90, Float/Vector Physical Reg.: 256, ARM v8 NEON (128-bit SIMD operations) \\
Memory System & L1: 32KB (I) 2-way + 32KB (D) 2-way, L2: 1MB 16-way, Main Memory: 8GB DDR3-1600 \\ \bottomrule
\end{tblr}
\end{table}

We execute Autoware's euclidean cluster algorithm in gem5, running in Full System mode (Ubuntu 18.04). We use gem5 fast-forwarding capabilities with KVM hardware virtualization~\cite{habib2008virtualization, sandberg2015FullSpeed} to reach the regions of the sub-sampled frames. For energy results, we model the CPU in McPAT~\cite{mcpat2009, mcpat2013} in a 32 nm technology, and use gem5 reported statistics to feed the McPAT power model. We estimate the area and power of the new \glspl{FU} (compression/decompression, and square of the differences with associated error) synthesizing Verilog descriptions on Synopsys Design Compiler~\cite{designCompiler} with a 14 nm technology~\cite{melikyan201814nm}. To unify results in a single technology we scale the baseline CPU data reported by McPAT using the methodology described by Stillmaker et al.~\cite{Stillmaker2017} (from 32 nm to 14 nm technology).

\subsection{Performance Analysis}\label{subsec:performance_analysis}

\begin{figure}[tb]
 \centering
 \includegraphics[width=0.48\textwidth]{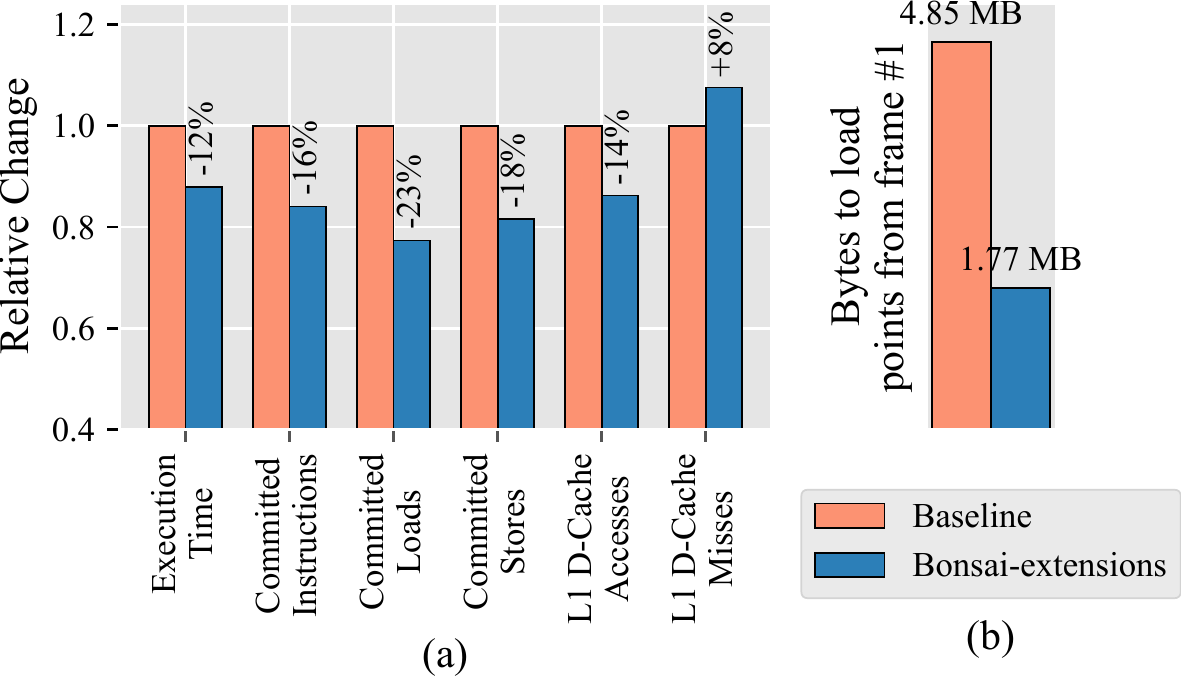}
 \caption{(a) Hardware metrics during the execution of the \textit{extract} kernel of euclidean clustering considering the baseline code and the proposed Bonsai-extensions. Average across all executed frames. (b) Number of loaded bytes to fetch points from the first frame of the data-set during \textit{radius search} (traversal).}
 \label{fig:hw_metrics}
 \vspace{-10pt}
 \end{figure}

Figure \ref{fig:hw_metrics}\bluecolor{a} presents key performance metrics for the execution of the \textit{extract} kernel of euclidean clustering, both for the baseline with and without the Bonsai-extensions. This is the main kernel of the algorithm and accounts for 90\% of its execution time (measured with Valgrind~\cite{ferrante2007valgrind}), and where both k-d tree build and search are performed. Since each metric has different scales, we normalized each of them w.r.t. the baseline code. We can see that the Bonsai-extensions reduce the number of memory instructions, by 23\% for loads and 18\% for stores. 

Figure \ref{fig:hw_metrics}\bluecolor{b} gives intuition for this improvement, depicting a great reduction in the number of required bytes to bring the \textit{points} from memory during the search on one frame. When we load compressed points using the Bonsai-extensions, we load a fraction (37\%) of the bytes we would normally need in the baseline code. Although this value is for the first frame of the data set, the behavior is similar across all frames.

This reduction in memory usage converts into several benefits. First, it decreases the number of committed instructions by 16\%, ultimately indicating that our Bonsai-extensions cut computation costs and increase efficiency on \textit{radius search} processing. Second, it reduces accesses to L1 D-cache by 14\%, making the application less memory-bound, which also increases efficiency in the use of the CPU. Third, due to both former reasons, it decreases the execution time of the \textit{extract} kernel by 12\%. Latency, as we further discuss, is a major concern for \gls{AD} algorithms~\cite{Lin2018ArchitecturalImplications}. Nevertheless, this is particularly significant when we observe that benefits come from the addition of only five new instructions to the \gls{ISA}.

\begin{figure}[bt]
 \centering
 \includegraphics[width=0.48\textwidth]{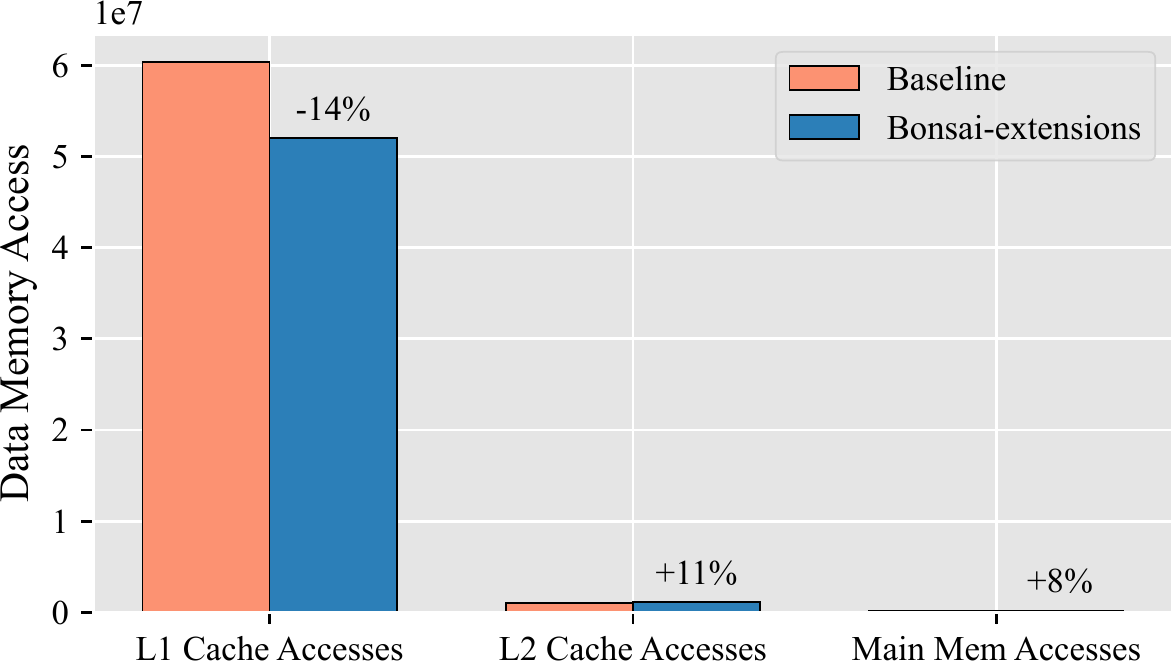}
 \caption{Accesses on different levels of the memory hierarchy.}
 \label{fig:mem_hierarchy}
\vspace{-10pt}
 \end{figure}

Figure \ref{fig:hw_metrics}\bluecolor{a} also indicates K-D Bonsai increases L1 D-cache misses. Although the Bonsai-extensions load compressed points from the \textit{cmprsd\_strct\_array}, which is contiguous in memory, it also accesses the original list of points when classification is inconclusive (white shell in Figure \ref{fig:radius_classification}). These infrequent accesses to another data structure are the main cause for misses in higher levels of the memory hierarchy. In absolute numbers, however, this is not a concern. Since the L1 cache is accessed $47\times$ more than L2 and 300$\times$ more than main memory we still see the benefits in execution time. Figure \ref{fig:mem_hierarchy} puts the number of memory accesses in perspective, according to the different memory hierarchy levels. This phenomenon highlights the importance of choosing the appropriate reduced \gls{FP} representation, as we discussed in Section \ref{subsec:compression_smaller_rep}, Table \ref{tab:smaller_rep_floating_point}, to minimize overheads of issuing 32-bit re-computation. In our experimentation, only 0.37\% of the classifications had to rely on the baseline computation. If we were not careful in selecting the representation, errors would not be as infrequent, and the K-D Bonsai benefits could be compromised.

\begin{figure}[bt]
 \centering
 \includegraphics[width=0.48\textwidth]{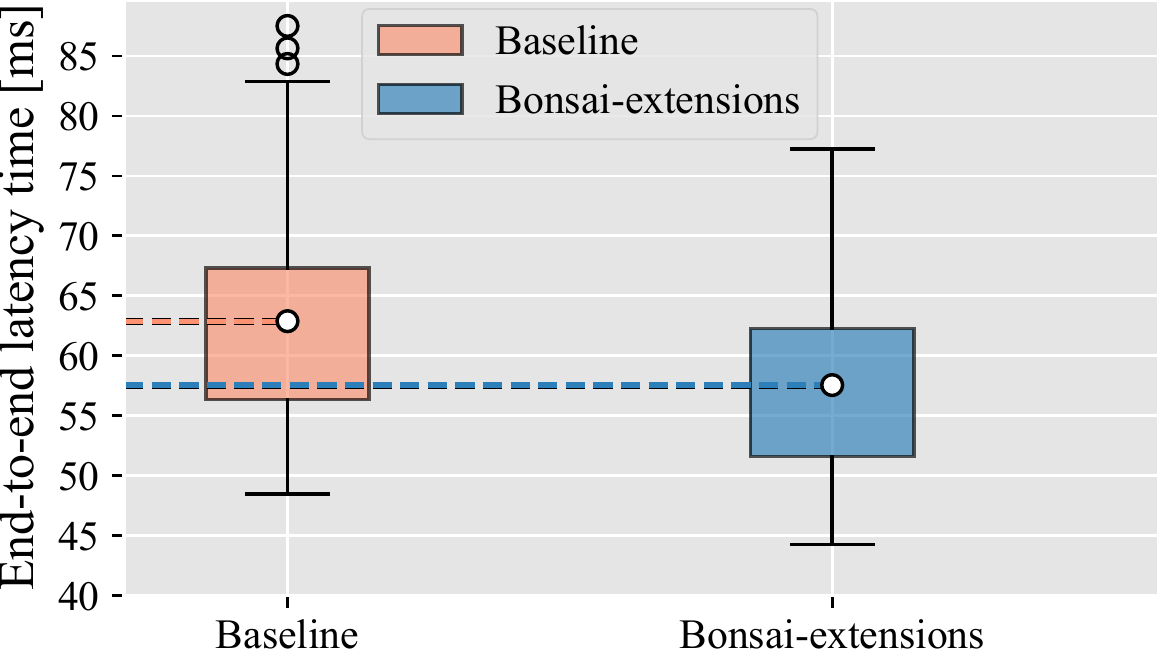}
 \caption{The distribution of the end-to-end latencies for the euclidean cluster algorithm. The dashed line indicates the mean value. Half the values are within the box limits.}
 \label{fig:latency_end_to_end}
 \vspace{-10pt}
 \end{figure}

Next, we evaluate end-to-end latency for euclidean cluster processing of frames. This is important because the \textit{extract} kernel, evaluated so far, is a subset of the algorithm's work. Other tasks such as point cloud pre-processing and labeling the points into their respective clusters must also be performed. Figure \ref{fig:latency_end_to_end} depicts two box plots with the distribution of the euclidean cluster end-to-end processing time for all sub-sample frames. As in any standard box plot, the boxes contain 50\% of the values. We indicate the mean value of each distribution (not the median, typical of box plots) with a white circle and auxiliary dashed lines. The use of Bonsai-extensions speeds up the average end-to-end latency by 9.26\%. In the context of \gls{AD}, reducing the end-to-end latency translates into reducing the reaction time of the vehicle, hence actuating faster, and increasing overall safety. At this point, we recall that K-D Bonsai benefits come with the same baseline accuracy (Section \ref{subsec:keep_accuracy}). Also, since the euclidean cluster is generally a perception bottleneck~\cite{Becker2020Demystifying, Kato2018AutowareOnBoards, zhao2020DrivingScenario} K-D Bonsai improvements are directly converted into overall \gls{AD} improvements.

Another important aspect for \gls{AD} algorithms is their end-to-end \textit{tail latency}. Different from the average, the tail latency assesses the performance of the algorithm in situations where computation takes the most (e.g., in the euclidean cluster, when point clouds have a higher number of points to be processed). K-D Bonsai again proves to be advantageous considering the 99th percentile tail latency, speeding it up by 12.19\%. Hence, K-D Bonsai improves performance when it is needed the most.

\subsection{Area and Power Analysis}\label{subsec:area_analysis}
Let us now examine the hardware costs of implementing our technique. Table~\ref{tab:area_power} presents area and power overheads introduced to support K-D Bonsai, according to the methodology explained in \ref{subsec:results_methodology}. Overall, the hardware to support the new instructions is simple, increasing area by 0.051 mm², which represents an increase of 0.36\% w.r.t the baseline. Likewise, supporting K-D Bonsai increases dynamic power by 24 mW (+1.29\% w.r.t the baseline). These results reinforce how non-intrusive our solution is. In the context of \gls{AD}, introducing minimal overheads in power and area are particularly important for meeting cooling constraints~\cite{Lin2018ArchitecturalImplications} and design of small autonomous vehicles (e.g.,  for delivery~\cite{nuro}), respectively.

% Please add the following required packages to your document preamble:
% \usepackage{booktabs}
% \usepackage{multirow}
\begin{table}[tb]
\centering
\caption{Area and Power for baseline CPU and K-D Bonsai}
\label{tab:area_power}
\begin{tabular}{@{}lllll@{}}
\toprule
\multicolumn{2}{l}{} & \multicolumn{1}{c}{Area (mm²)} & \multicolumn{1}{c}{\begin{tabular}[c]{@{}c@{}}Dynamic \\ Power (W)\end{tabular}} & \multicolumn{1}{c}{\begin{tabular}[c]{@{}c@{}}Static \\ Power (W)\end{tabular}} \\ \midrule
\multicolumn{2}{l}{Processor (L2 included)} & 14.26 & 1.86 & 1.15 \\ \midrule
\multirow{3}{*}{\begin{tabular}[c]{@{}l@{}}K-D \\ Bonsai\end{tabular}} & \begin{tabular}[c]{@{}l@{}}Compression\\ Decompression\\ FU\end{tabular} & 0.0191 & 0.0095 & 6.29E-06 \\ \cmidrule(l){3-5} 
 & 4x (A-B’)² FU & 0.0320 & 0.0144 & 4.55E-06 \\ \cmidrule(l){3-5} 
 & Total & 0.0511 & 0.0240 & 1.08E-05 \\ \midrule
\multicolumn{2}{l}{Relative change} & 0.36\% & 1.29\% & 0.001\% \\ \bottomrule
\end{tabular}
\end{table}

\subsection{Energy Analysis}\label{subsec:energy_analysis}

\begin{figure}[tb]
 \centering
 \includegraphics[width=0.48\textwidth]{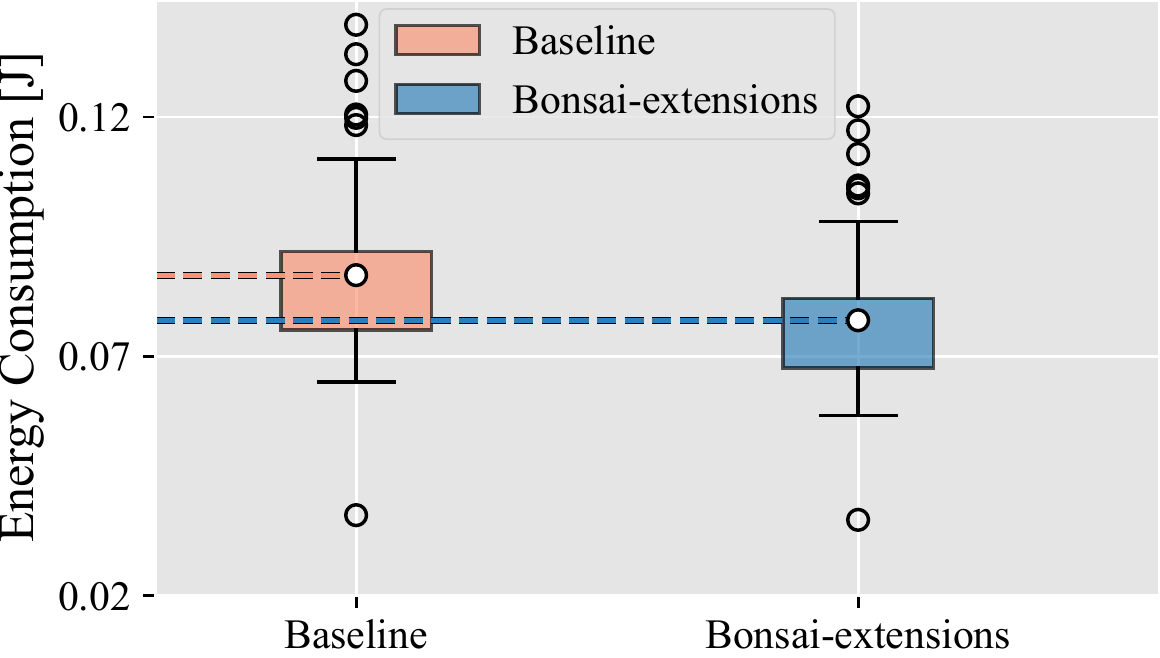}
 \caption{The distribution of the energy consumption for the \textit{extract} kernel in the euclidean cluster algorithm. The dashed line indicates the mean value. Half the values are within the box limits.}
 \label{fig:energy_kernel}
 \vspace{-10pt}
 \end{figure}

Finally, we go through K-D Bonsai energy consumption results in the \textit{extract} kernel of the euclidean cluster. Figure \ref{fig:energy_kernel} depicts a box-plot (in the same fashion we did for end-to-end latency, Figure \ref{fig:latency_end_to_end}). The reduction in energy consumption is driven by a reduction in execution time, number of instructions and number of memory accesses, which pays off the small increase in dynamic power (Table \ref{tab:area_power}). On average, the use of Bonsai-extensions reduces energy consumption by 10.84\%. K-D Bonsai successfully improves energy efficiency, which is a concern on \gls{AV} so the computational platform does not reduce driving range~\cite{Lin2018ArchitecturalImplications} (e.g., on battery-powered vehicles).

\section{Related Work} \label{sec:related_work}

Recent advances in \gls{LiDAR} technology and \gls{AV} motivated previous works on improving point cloud processing. Heinzle et al. propose specific hardware to improve radius and nearest-neighbor search in k-d tree point clouds~\cite{Heinzle2008}. Their main idea is to search slightly more points than necessary (e.g., asking for a larger radius), and use the extended result-set for subsequent, spatially closed queries. They implement it in an FPGA and, compared to a CPU, it improves query throughput by 68\%, \textit{ignoring} CPU-FPGA transfer costs. The work claims the used platform lacks an efficient CPU-FPGA interface, resulting in \textit{half} of the baseline CPU performance if communication costs are taken into account. This highlights an important challenge for accelerator integration. Since K-D Bonsai is implemented as part of the CPU, there is no overhead to transfer data in and out of the core. Also, while their work focuses on speeding up the traversal, K-D Bonsai improves \textit{leaf processing}, reducing the total number of loads to bring the data (via (de)compression in the CPU) after the traversal is performed, hence being orthogonal to their technique.

A more recent work introduces Tigris~\cite{Xu2019}, an accelerator to speed up radius and nearest-neighbor search for point cloud registration (a major application for k-d tree search, see Section \ref{sec:introduction}). Tigris divides traversal and leaf processing in a front-end/back-end fashion. Multiple queries are traversed in parallel, offloading leaf data to the back-end, where multiple \glspl{FU} will perform distance checks. To exploit higher performance, Tigris also has a scheme to search on previously obtained result-sets, causing their search to be \textit{approximate}. The accelerator improves end-to-end latency for registration w.r.t a CPU by 86.6\% but requires a total area of 15.57 mm$^2$ (more than our Baseline CPU, see Table \ref{tab:area_power}). The work does not report the cost to offload queries to the accelerator.
QuickNN~\cite{pinkham2020QuickNN} also accelerates nearest-neighbor search on k-d tree-based point clouds. In their target application, point cloud frames are used as references for new frames. They exploit this behavior in the accelerator architecture, overlapping execution and sharing data of tree build and tree traversal. Moreover, they propose a gather-read and gather-write cache, coalescing accesses to off-chip memory. Like Tigris, their work processes multiple queries in parallel and performs an approximate search. Accelerators for nearest-neighbor search on high-dimensional spaces were also proposed~\cite{lee2022ANNA,abselhadi2019AcceleratedANNSearch}, but the problem properties and requirements differ from 3D data used by \gls{AD}.

%Mesorasi and pointacc
The Mesorasi~\cite{Feng2020} accelerator for point cloud based \glspl{CNN} (e.g. PointNet++~\cite{Li2017PointNet++}) proposes delayed-aggregation, allowing neighbor search and feature computation to be overlapped in time, hiding latency. In their proposal, point cloud search time stays roughly the same, and most of the benefits come from faster feature computation. A more recent work~\cite{Feng2022Crescent} improves over Mesorasi by limiting the backtracking step of the k-d tree search to a sub-tree at the cost of accuracy. Additionally, in case of bank conflicts their solution re-uses similar points or completely ignores traversal paths if necessary. This approximate scheme has most of its accuracy corrected during training, restricting it to machine learning scenarios. Similarly to Mesorasi, PointAcc~\cite{Lin2021PointAcc} also accelerates point cloud-based \glspl{CNN}, proposing a ranking-based generic accelerator unit. They also increase the number of mapping operations over Mesorasi~\cite{Feng2020}, e.g., supporting \textit{radius search}, and farthest point sampling to fetch inputs in the point cloud.

Some works exploited GPUs to improve point cloud processing. The Buffer k-d tree~\cite{gieseke2014bufferkdtree} proposes nearest-neighbor search using a buffer to delay the processing of queries of the same leaf until enough work is gathered. RTNN~\cite{Zhu2022RTNN} proposes to formulate neighbor search into a ray tracing problem. It then exploits contemporary ray tracing hardware in GPUs to improve the search. The work, however, shows to be effective on point clouds orders of magnitude (hundreds of thousands to millions of points) bigger than those generally processed in one \gls{LiDAR} frame (thousands of points), as data transfer overhead shows to be increasingly relevant as the point cloud size decreases. Nguyen et al.~\cite{nguyen2020FastEuclidean} focuses on the software perspective, implementing the euclidean cluster task with different data structures and observing their efficiency in the GPU hardware. Nonetheless, evaluation of Autoware.ai algorithms had shown that using the GPU for the euclidean clustering performs similarly to an \gls{OoO} CPU due to the GPU communication overheads~\cite{Kato2018AutowareOnBoards}. Indeed, Autoware.ai uses the CPU instead of the GPU to run the euclidean cluster by default~\cite{autowareGitHub}. The GPU is reserved for image-based object detection \glspl{CNN} where its benefits are much less debatable~\cite{Kato2018AutowareOnBoards}. At the same time, adding a GPU just for point cloud processing incurs orders of magnitude more area and power overheads than \textit{K-D Bonsai}, both critical aspects to be minimized in \glspl{AV}~\cite{Lin2018ArchitecturalImplications}.

In summary, the majority of previous works propose the use of accelerators, implying communication overheads, high area costs, and lack of programmability. Moreover, many works introduce approximation to achieve effective solutions. On the other hand, \textit{K-D Bonsai} proposes a small set of new instructions implemented inside a CPU, while also guaranteeing baseline accuracy. This fundamental difference places K-D Bonsai as a programmable and easy-to-be-adopted solution in nowadays systems. Our solution is modest performance-wise, but orders of magnitude cheaper regarding area and power. Nevertheless, our (de)compression scheme aims for better data fetching for leaf processing, an orthogonal concept unexploited by previous works. The techniques presented here, we recall, are also applicable in platforms other than a CPU, as it only depends on the algorithm (\textit{radius search}) and the input (\textit{point clouds}).

\section{Conclusions} \label{sec:conclusions}

In this work, we proposed K-D Bonsai, a novel approach to improve leaf processing during k-d tree radius search, a key operation in modern point cloud processing algorithms for \glspl{AV}. K-D Bonsai reduces memory operations with a (de)compression scheme that takes advantage of value similarity and precision reduction tolerance in the points of k-d tree leaves, without harming baseline accuracy. Differently from previous works that rely on out-of-core accelerators, K-D Bonsai is implemented in the form of \gls{ISA}-extensions (Bonsai-extensions) in an \gls{OoO} CPU, and validated with state-of-the-art software for \gls{AV} in a cycle-accurate simulator in full-system mode. Our solution, K-D Bonsai, proved to be very efficient in reducing both end-to-end latency and energy consumption while incurring minimal overheads in area and power. Besides, it requires only incremental hardware modifications on commodity CPUs while being simple to be used by the programmer (setting a flag in \gls{PCL}), hence being a hands-down solution for next-generation \gls{AV} systems.

\section{Acknowledgements} \label{sec:ack}
This work has been supported by the CoCoUnit ERC Advanced Grant of the EU’s Horizon 2020 program (grant No 833057), the Spanish State Research Agency (MCIN/AEI) under grant PID2020-113172RB-I00, the ICREA Academia program, and a FPI-UPC 2020 grant.

\balance

%%%%%%% -- PAPER CONTENT ENDS -- %%%%%%%%

%%%%%%%%% -- BIB STYLE AND FILE -- %%%%%%%%
\bibliographystyle{IEEEtranS}
\bibliography{refs}
%%%%%%%%%%%%%%%%%%%%%%%%%%%%%%%%%%%%

\end{document}